\documentclass[onecollarge]{svjour2}
\smartqed
\usepackage{graphicx,amssymb,amsmath}

\usepackage{color}

\newcommand{\be}{\begin {equation}}
\newcommand{\ee}{\end{equation}}
\newcommand{\bi}{\begin{itemize}}
\newcommand{\ei}{\end{itemize}}
\newcommand{\bea}{\begin {eqnarray}}
\newcommand{\eea}{\end{eqnarray}}
\newcommand{\bra}[1]{\langle\,#1\,|}          
\newcommand{\ket}[1]{|\,#1\,\rangle}          

\newcommand{\LCm}{{\scriptscriptstyle -}} 
\newcommand{\LCp}{{\scriptscriptstyle +}}

\journalname{Few-Body Systems}
\begin{document}


\title{Nonperturbative Quantum Field Evolution}

\author{Xingbo~Zhao \and 
        Anton~Ilderton  \and
        Pieter~Maris  \and
        James~P.~Vary}

\institute{X.~Zhao, P.~Maris, J.~P.~Vary \at
Department of Physics and Astronomy, Iowa State University, Ames, IA 50011, USA, \email{xbzhao@iastate.edu, pmaris@iastate.edu, jvary@iastate.edu}
          \and
           A.~Ilderton \at
         Department of Applied Physics, Chalmers University of Technology, SE-412 96 Gothenburg, Sweden, \email{anton.ilderton@chalmers.se}
}

%





%

\date{\today}
\maketitle 
\begin{abstract}

We introduce a nonperturbative, first-principles approach to time-dependent problems in quantum field theory. In this approach, the time-evolution of quantum field configurations is calculated in real time and at the amplitude level. This method is particularly suitable for treating systems interacting with a time-dependent background field. As a test problem, we apply this approach to QED and study electron acceleration and the associated photon emission in a time- and space-dependent electromagnetic background field. 
%
\end{abstract}


%
\section{Introduction}

Light-front dynamics offers advantages for nonperturbative calculations in quantum field theory (QFT) thanks to the fact that the vacuum structure is simple and the Lorentz boosts are kinematic~\cite{Brodsky:1997de}. Based on light-front dynamics and the Hamiltonian formalism, the Basis Light-front Quantization (BLFQ) approach~\cite{Vary:2009gt,Honkanen:2010rc} was previously constructed for stationary problems in QFT. In BLFQ, the bound state problem is treated as the eigenvalue problem of the light-front Hamiltonian of the system. In order to explore the dynamics of QFT, we extend BLFQ to the time-dependent regime and construct the time-dependent BLFQ (tBLFQ) approach~\cite{Zhao:2013cma,Zhao:2013jia}. In tBLFQ, we simulate time-dependent processes by computing time-evolution of quantum field configurations at the amplitude level.

One of the typical applications of tBLFQ is scattering processes, and tBLFQ is particularly suitable for dealing with scattering in the presence of a time-dependent background field. In this paper, we introduce tBLFQ through the process of photon emission in a time-dependent background field, modeling an intense laser~\cite{Jaeckel:2010ni,DiPiazza:2011tq,Heinzl:2011ur}. In this process, often referred to as ``nonlinear Compton scattering''(nCs) ~\cite{Nikishov:1963}, an electron is accelerated by the background and a photon is emitted into the final state. While the standard perturbative approach to QED can be extended to allow a treatment of arbitrarily strong backgrounds for very simple background field models, we will here treat all interactions nonperturbatively (see also e.g.~\cite{Hebenstreit:2011wk} for other nonperturbative approaches to QED in strong backgrounds.)

This paper is organized as follows: in Sec.~\ref{sec:background}, we present the background field profile used in this calculation; in Sec.~\ref{sec:tBLFQ} we introduce the formalism of tBLFQ; in Sec.~\ref{sec:BLFQ} we briefly review BLFQ which is employed to construct the basis states for time-evolution in tBLFQ; in Sec.~\ref{sec:results} we present the numerical results for a sample calculation of the nCs. Finally we conclude and provide our outlook for future work in Sec.~\ref{sec:conclusion}.

\section{Background Field Profile}
\label{sec:background}
We model the background as a classical field. We neglect the quantum fluctuation as well as back reaction on the background field. Specifically, we choose an electric field in the 3-direction, with profile,
 \begin{align}
 E^3(x^+,x^-)=-E^3_0 \sin{(l_\LCm x^\LCm)} \sin{(l_\LCp x^\LCp)} \Theta(x^\LCp)\Theta(\Delta x^\LCp-x^\LCp)\  ,
 \label{eq:laser_profile_electric_field}
 \end{align}
where $E^3_0$ is the peak amplitude, $l_\LCm$ ($l_\LCp$) is the frequency in the longitudinal (light-front time) direction. The theta functions impose a finite light-front time duration, $\Delta x^\LCp$, on the field. In the light-cone gauge ($A^+$=0), which we adopt throughout this paper, an appropriate gauge potential is
\begin{align}
\mathcal{A}^-(x^+,x^-)=\frac{E^3_0}{l_\LCm} \cos{(l_\LCm x^\LCm)} \sin{(l_\LCp x^\LCp)} \Theta(x^\LCp)\Theta(\Delta x^\LCp-x^\LCp)\ .
\label{eq:laser_profile}
\end{align}
As the goal of this paper is to demonstrate tBLFQ, rather than to provide precise phenomenological predictions, it is not an issue that our chosen background profile does not satisfy Maxwell's equation in vacuum.

\section{Quantum Evolution}
\label{sec:tBLFQ}

The tBLFQ adopts the Hamiltonian formalism of QFT. In tBLFQ, the evolution of quantum field configurations is calculated through solving the time-dependence of the system's total amplitude (expressed as a superposition of configurations) generated by the light-front Hamiltonian. Probability amplitudes for a specific final state may be obtained by projecting the evolved state onto that final state.

In the example of nCs process, the Hamiltonian $P^-$ consists of the {\it full}, interacting, light-front Hamiltonian of QED, $P^-_\text{QED}$ along with the external field interactions $V(x^\LCp)$, namely,
\begin{align}
	\label{H_interact}
	P^-(x^+)=P^-_\text{QED}+V(x^+) \;.
\end{align}

Over time, both the QED Hamiltonian $P^-_\text{QED}$ and the external field interaction $V(x^\LCp)$ may induce transitions on the quantum field amplitudes.  In the nCs process, the explicit time-dependence, however, only enters through $V(x^\LCp)$. Therefore, the time-evolution problem can be most easily solved in an interaction picture, where the time-independent light-front QED Hamiltonian $P^-_\text{QED}$ serves as the ``main'' part of the Hamiltonian (usually the ``free" part, but we write ``main" to underline that we are not dealing with the usual ``free" + ``interacting" split) and the time-dependent external field interaction $V(x^\LCp)$ as the ``interacting'' part. In this interaction picture, the quantum field amplitude evolves according to,
\be\label{Schro-int}
	i\frac{\partial}{\partial x^+}\ket{\psi;x^+}_I= \frac{1}{2}V_I(x^+)\ket{\psi;x^+}_I \;,
\ee
where $\ket{\psi;x^+}_I=e^{iP^-_\text{QED}x^+/2}\ket{\psi;x^+}$ is the quantum field amplitude in the interaction picture, and the ``interaction Hamiltonian in the interaction picture'' $V_I$ evolves in time according to
\be\label{V-int}
	V_I(x^+) = e^{\tfrac{i}{2}P^-_\text{QED}x^+}V(x^+)e^{-\tfrac{i}{2}P^-_\text{QED}x^+} \;.
\ee
The solution to~(\ref{Schro-int}) can be formally written in terms of a time-ordered ($ \mathcal{T}_+$) series, as:
\begin{align}
	\label{i_evolve}
	\ket{\psi;x^+}_I  &= \mathcal{T}_\LCp e^{-\tfrac{i}{2}\int\limits_{x^+_0}^{x^+} V_I\, dx^+}\ket{\psi;x^+_0}_I \;,
\end{align}
where $\ket{\psi;x^+_0}_I$ is the initial quantum field amplitude at light-front time $x^+_0$. To make the numerical calculation feasible, we approximate Eq.~(\ref{i_evolve}) by decomposing the time-evolution operator into small but finite steps, with the step size $\delta x^+$. Each step is implemented as a matrix-vector multiplication acting on the state vector of the previous step, as
\be
\label{eq:sol_wave-eq_i_discrete}
\mathcal{T}_\LCp e^{-\frac{i}{2}\int\limits_{x^+_0}^{x^+} V_I\, dx^+}\ket{\psi(x^+_0)}_I\rightarrow \big[1-\tfrac{i}{2}V_I(x^+_{n})\delta x^+\big] \cdots \big[1-\tfrac{i}{2}V_I(x^+_{1})\delta x^+\big]\ket{\psi(x^+_0)}_I \;.
\ee
In the nCs process, the initial state, $\ket{\psi(x^+_0)}$, corresponds to a single physical electron. 

Next, we need to specify a basis to represent the quantum field amplitudes $\ket{\psi;x^+}_I$ as well as the external field interaction $V_I(x^+)$. The most convenient basis is made of the eigenstates of the time-independent part of the Hamiltonian, that is, the light-front QED Hamiltonian, $P^-_\text{QED}$. We denote such a basis as $\ket{\beta}$, which can be found by solving the eigenvalue problem for $P^-_\text{QED}$ in BLFQ~\cite{Vary:2009gt,Honkanen:2010rc}. More details on this step will be shown in the next section. In $\ket{\beta}$, the initial state in the nCs process -- a physical electron -- can be directly expressed, for a simple example, as an eigenstate of $P^-_\text{QED}$.

\section{Basis Construction}
\label{sec:BLFQ}

In this section we present a brief review of BLFQ~\cite{Vary:2009gt,Honkanen:2010rc} and explain the procedure of constructing the tBLFQ basis $\ket{\beta}$ out of a chosen initial basis. For more details, see Ref.~\cite{Zhao:2013cma}. 

The central idea in BLFQ is to solve the eigenvalue problem of QFT in an efficient basis to mitigate the computational burden. The chosen initial basis used in BLFQ, denoted as $\ket{\alpha}$, explicitly exploits the following symmetries of the light-front Hamiltonian: 1) Translational symmetry in the $x^-$ direction; 2) Rotational symmetry in the transverse plane; 3) Lepton number conservation. Consequently, such a basis divides the eigenspace of the Hamiltonian into subspaces labelled by the conserved charges of these three symmetries, which are the longitudinal momentum $P^\LCp$, longitudinal projection of angular momentum $M_j$, and the charge, or net fermion number $N_f$, respectively. In this paper, the longitudinal direction ($x^-$) is compactified to a circle of length $2L$, with the (anti-)periodic boundary condition imposed for bosons (fermions). Thus, the longitudinal momentum is discretized: $P^\LCp=2\pi K/L$, where an integer (or half-integer) $K$ is used to label the longitudinal momentum. In tBLFQ, we refer to each subspace designated by the group of quantum numbers \{$K$, $M_j$, $N_f$\} as a {\it segment}.

In general, the background field can exchange longitudinal momentum as well as longitudinal projection of angular momentum with the system. Therefore, one needs to solve the eigenvalue problem for the tBLFQ basis states $\ket{\beta}$ in multiple segments. In this paper, our chosen background field profile~(\ref{eq:laser_profile}) only transfers longitudinal momentum into the system and does not make transverse excitations. Thus, we only need to work with segments with $M_j$=1/2, $N_f$=1 and different $K$'s.

In each segment, the BLFQ basis $\ket{\alpha}$ is constructed in terms of a Fock-sector expansion. Each Fock particle is represented in terms of longitudinal, transverse and helicity degrees of freedom. For longitudinal degrees of freedom, a plane-wave basis is used with the wave number $k$ as the quantum number. For transverse degrees of freedom, the eigenstates of a 2D-harmonic oscillator (2D-HO) are employed as the basis states with the radial quantum number $n$ and the angular quantum number $m$. The helicity degree of freedom is labelled by the helicity quantum number $\lambda$. 

In order to keep the basis size finite, truncations are made both on the Fock-sector level and inside each Fock sector. In this work, we make the lowest nontrivial Fock-sector truncation by retaining only the single electron sector $\ket{e}$ and the electron-photon sector $\ket{e\gamma}$. Inside each Fock sector, we adopt the following ``$N_{\max}$'' truncation of the transverse degrees of freedom:  We define the following sum over the 2D-HO quantum numbers of all Fock particles in a basis state, $N_\alpha=\sum_i 2n_i+|m_i|+1$ and then retain only those basis states whose $N_\alpha$ not greater than a chosen cutoff $N_{\max}$. The truncation on the longitudinal degrees of freedom is implicit: the fact that the wave number $k$ for each Fock particle is an integer or half-integer and they sum up to the segment label $K$ limits the number of possible longitudinal momentum partitions in each segment.

With the BLFQ basis $\ket{\alpha}$ constructed, we diagonalize the light-front QED Hamiltonian $P^-_\text{QED}$ segment by segment and obtain the eigenstates $\ket{\beta}$, which serve as the basis for the time-evolution~(\ref{eq:sol_wave-eq_i_discrete}). The nonperturbative renormalization procedure in BLFQ follows the guidance of a sector-dependent scheme~\cite{Karmanov:2008br,Zhao:2013xx}. In this paper, as a demonstration of tBLFQ, we neglect renormalization for simplicity. The resulting ground states (with the lowest invariant mass in each segment) are interpreted as the physical electron states and the excited states are interpreted as the electron-photon scattering states.

\section{Numerical Results}
\label{sec:results}

In this section we present numerical results for a sample calculation of the nCs process. A basis consisting of three segments with $K{=}\{K_i, K_i{+}k_\text{las}, K_i{+}2k_\text{las}\}$ is chosen for this calculation. 
The initial state for our process is a single electron (ground) state in the $K{=}K_i$ segment. This basis allows for the ground state to be excited twice by the background field (from the segment with $K$=$K_i$ through to the segment with $K_i$+2$k_\text{las}$). In this calculation, we take $K_i{=}1.5$ and $N_\text{max}{=}8$, with $a_0$=0.1, $k_\text{las}$=2, $L{=}2\pi$\,MeV$^{-1}$, $l_\LCm$=$\frac{\pi}{L}k_\text{las}{=}1$\,MeV and $l_\LCp{=}\frac{2\pi}{50}$\,MeV. 

\begin{figure}[!tbh]
\centering
\includegraphics[width=0.65\textwidth]{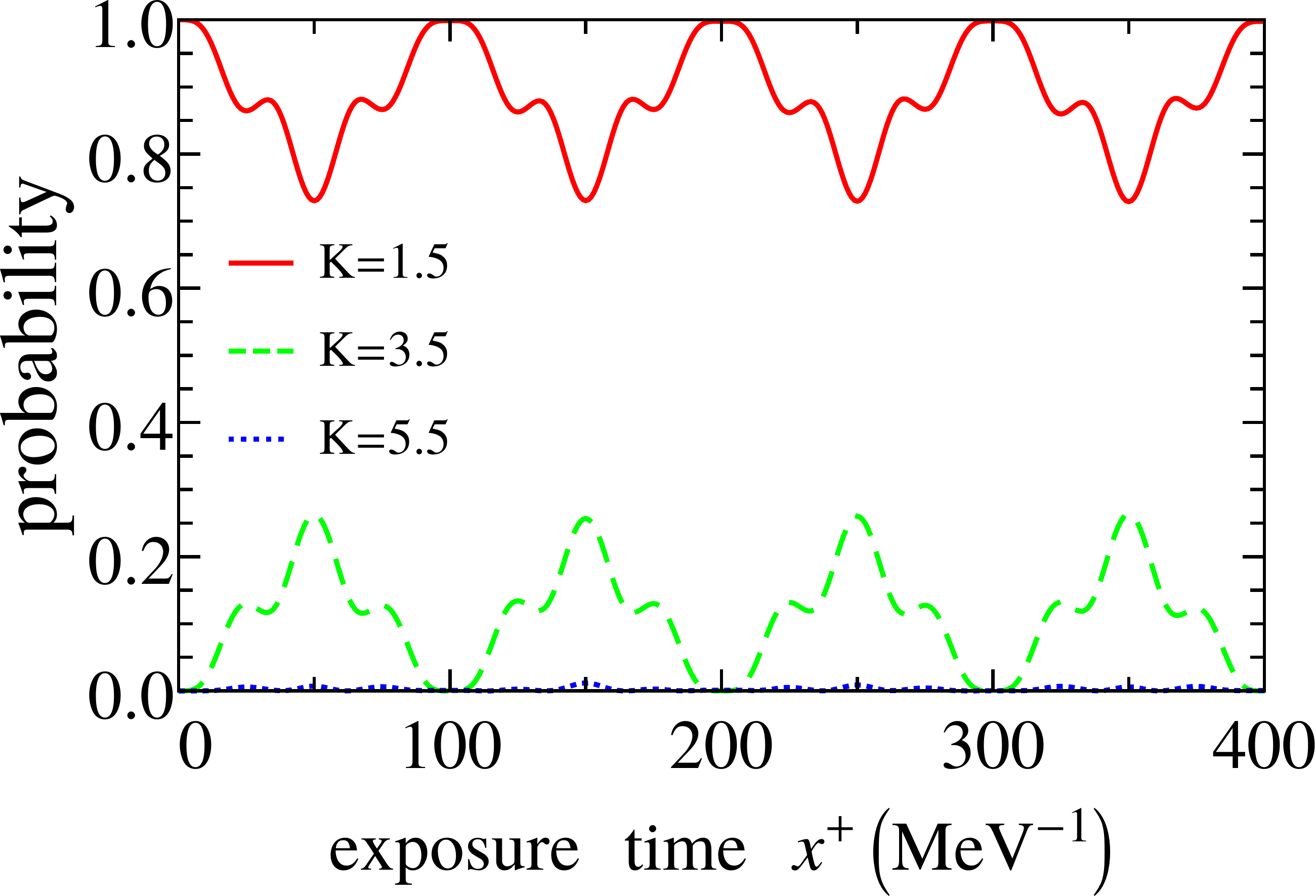}
\caption{\label{fig:ground_state_evol_np} (Color online) Time evolution of the probabilities of finding the three single physical electron states in the segments $K{=}1.5, 3.5, 5.5$.} 
\end{figure}

First, we display the time-evolution of the three physical electron states in segments with $K$=1.5, 3.5, 5.5 in Fig.~\ref{fig:ground_state_evol_np}. The probabilities of finding these three states oscillate with time in a (approximate) periodic pattern. This pattern reflects the fact that the physical electron is accelerated and decelerated in the longitudinal direction by the background field. The oscillating period $\sim$100Mev$^{-1}$ is determined by both the frequency of the background field in the longitudinal direction, $l_\LCm$, and that in the light-front time, $l_\LCp$.

\begin{figure}[!tbh]
\centering
\includegraphics[width=0.65\textwidth]{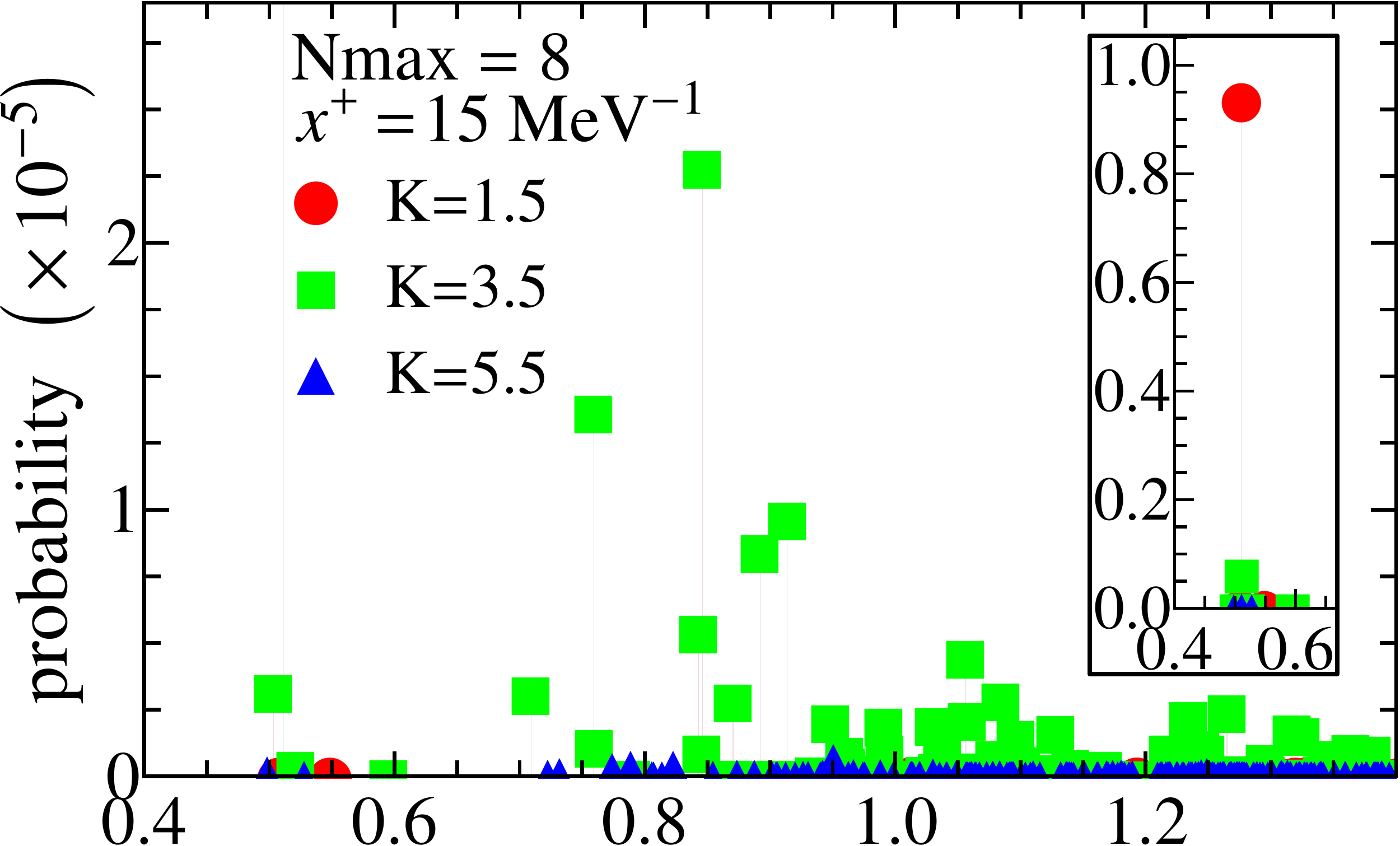}
\includegraphics[width=0.65\textwidth]{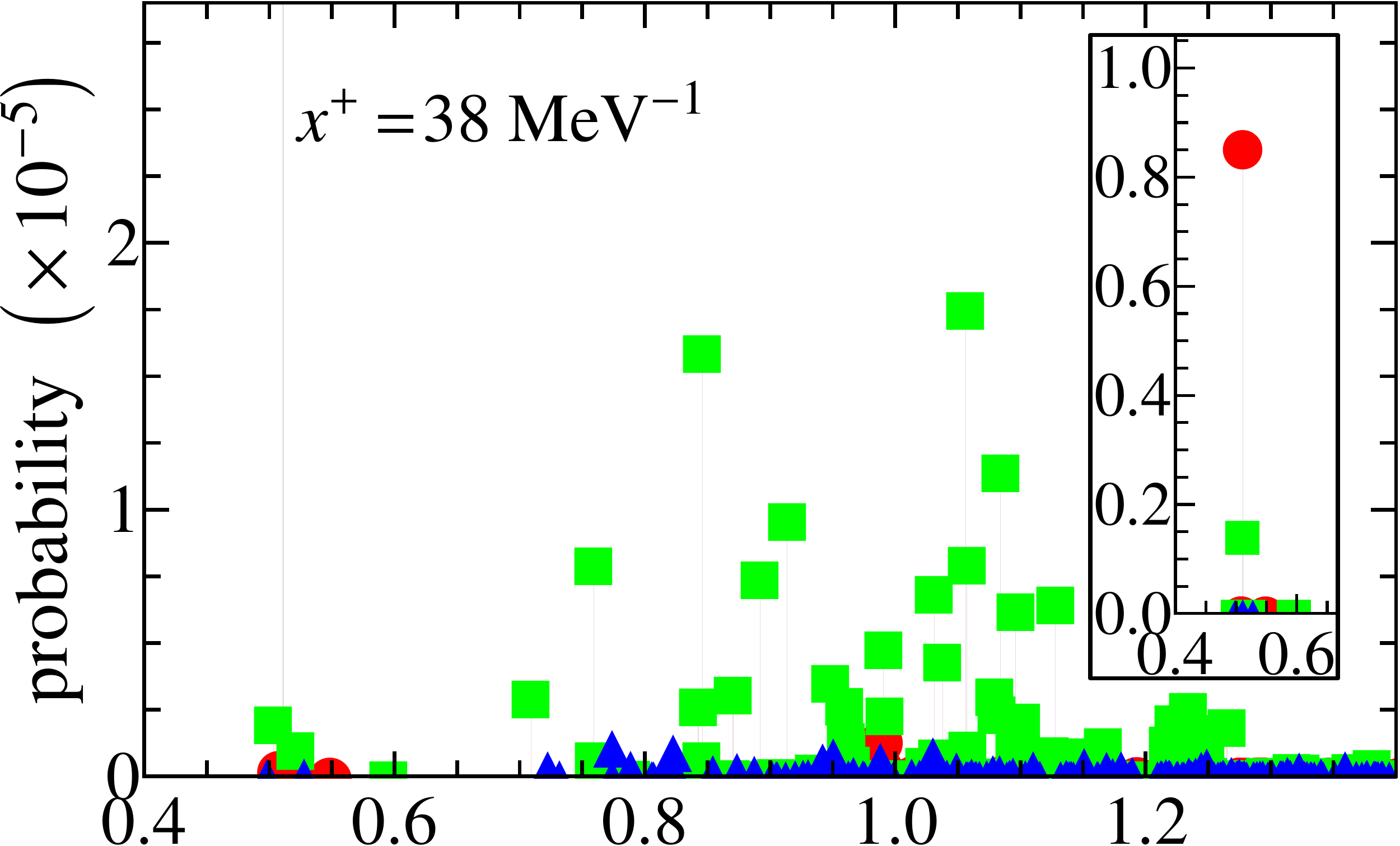}
\includegraphics[width=0.667\textwidth]{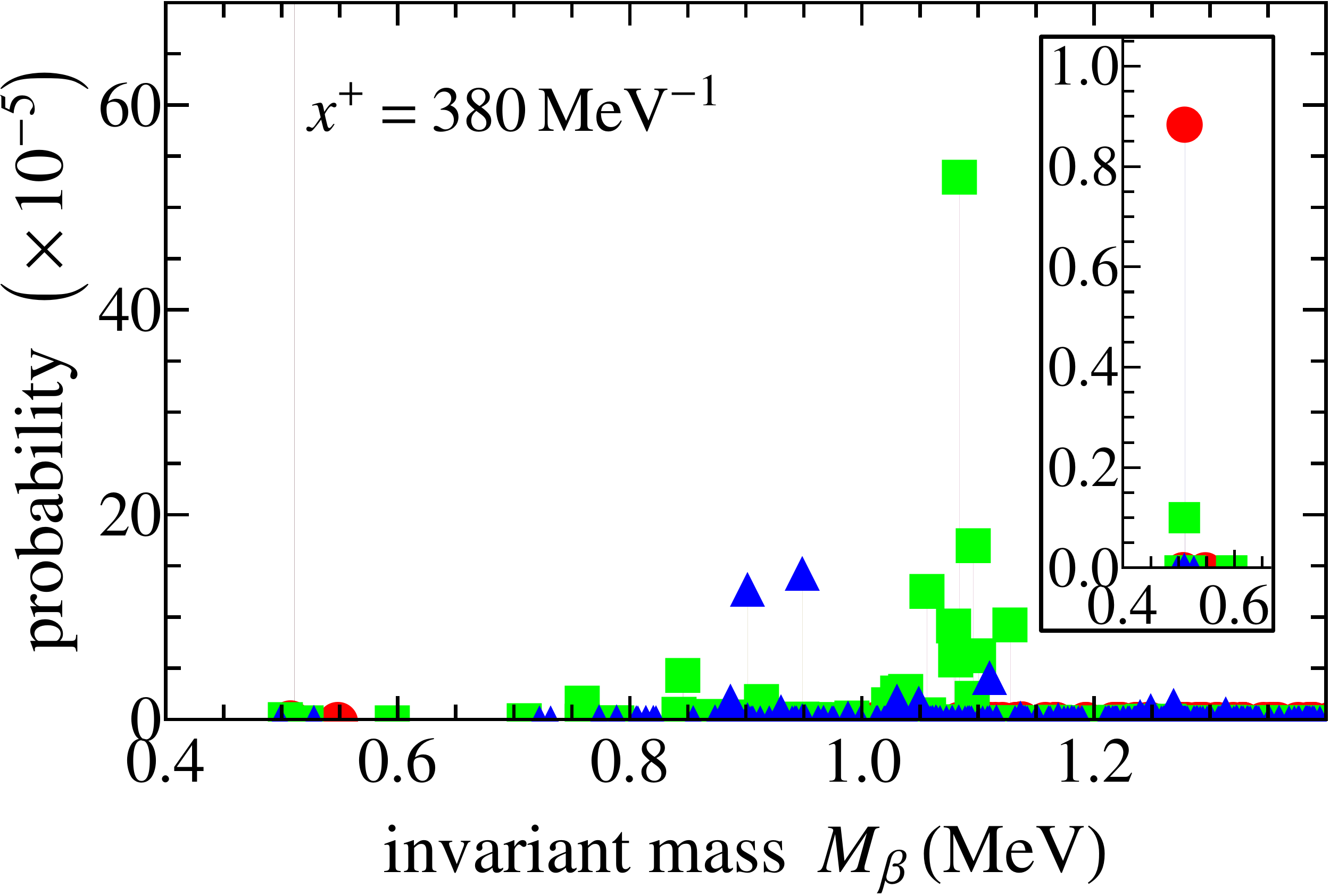}
\hspace{1.5mm}
\caption{\label{fig:state_evol_np} (Color online) Snapshots of the system at (from top to bottom) $x^+= 15,38,380$\,MeV$^{-1}$, with the background field switched on (off) at $x^+$=0($x^+$=400MeV$^{-1}$). Each dot represents a tBLFQ basis state $\ket{\beta}$. Horizontal axis: the invariant mass, $M_{\beta}$, of the state $\ket{\beta}$. Vertical axis: the probability of finding $\ket{\beta}$ in units of $10^{-5}$. The inset panels show, at normal scale, the (much larger) probabilities of finding the three single physical electron states in the segments with $K{=}1.5, 3.5, 5.5$.} 
\end{figure}

The periodicity in Fig.~\ref{fig:ground_state_evol_np} is only approximate, because as the electron accelerates, it radiates a photon, which takes away energy from the electron. To illustrate the photon emission process, we present several selected snapshots of the system in Fig.~\ref{fig:state_evol_np}, at increasing (top to bottom) light-front times. As time evolves, the electron-photon states (those with the invariant mass $M_{\beta}>$ 0.511\,MeV) are gradually populated. This reflects the fact that the photon radiation occurs along with the acceleration (and deceleration) of the single electron states.
%

\begin{figure}[t]
\centering
\includegraphics[width=0.68\textwidth]{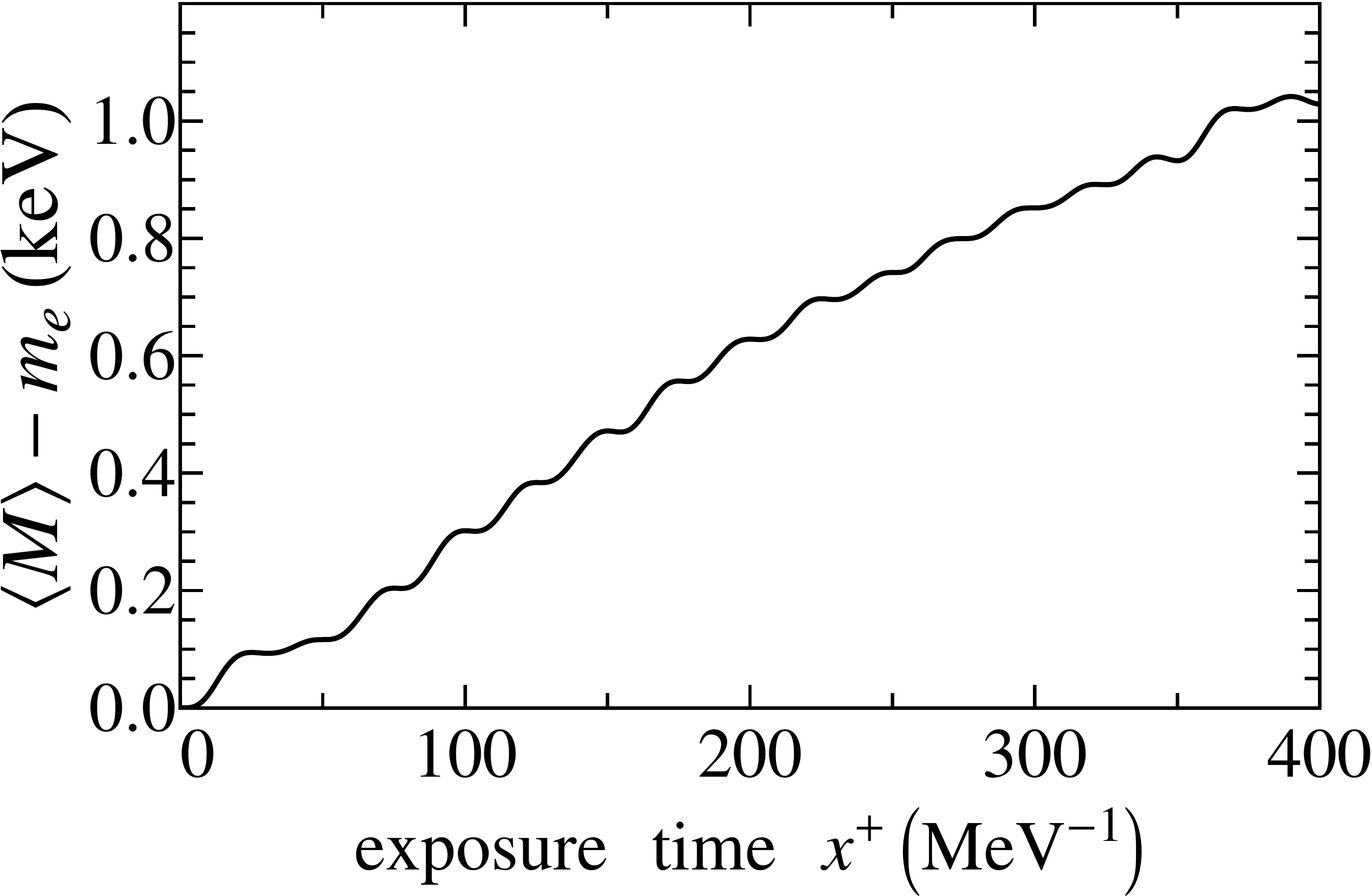}
\caption{\label{fig:invmass_evol_np} (Color online) The average invariant mass of the electron-photon system evolves as a function of exposure time. Horizontal axis: the exposure time. Vertical axis: the difference between the average invariant mass of the electron-photon system and that of a single physical electron.}
\end{figure}

During the evolution, the amplitude of the system at any intermediate time $\ket{\psi;x^\LCp}$ is available, which encodes all the information of the system and can be used to construct observables. As an example, Fig.~\ref{fig:invmass_evol_np} shows the evolution of the average invariant mass $\langle M(x^\LCp)\rangle\equiv\sum_\beta M_\beta\bra{\beta}\psi;x^\LCp\rangle^2$ of the system as a function of exposure time. The approximately linear increase of the invariant mass reflects the fact that energy is pumped into the system by the background field.

\section{Conclusion and Outlook}
\label{sec:conclusion}
In this paper, we have introduced a first-principles nonperturbative framework for time-dependent problems in quantum field theory. Named as ``time-dependent BLFQ" (tBLFQ), this approach takes the light-front Hamiltonian of the system, and an initial state as input. It solves for the evolved quantum field configurations at the amplitude level. Basis truncation and time-step discretization are the only two approximations.

We have illustrated tBLFQ through an application to the nonlinear Compton scattering process in QED where an electron is accelerated by a time-dependent background field and emits a photon. The numerical results reveal a coherent superposition of electron acceleration and photon emission processes.





We envision that another application of tBLFQ is the energy loss of the produced quark and gluon jets in relativistic heavy-ion collisions, where the evolving medium formed by the colliding nuclei can be modeled as a time-dependent background field.

We acknowledge valuable discussions with K. Tuchin, H. Honkanen, S. J. Brodsky, P. Hoyer, P. Wiecki and Y. Li. This work was supported in part by the Department of Energy under Grant Nos. DE-FG02-87ER40371 and DESC0008485 (SciDAC-3/NUCLEI) and by the National Science Foundation under Grant No. PHY-0904782. A.~I.\ is supported by the Swedish Research Council, contract 2011-4221.


\end{document}